\documentclass[numreferences]{kluwer}
\usepackage[english]{babel}
\usepackage[dvips]{graphicx}

\newcommand{\beq}{\begin{equation}}
\newcommand{\eeq}{\end{equation}}
\newcommand{\bea}{\begin{eqnarray}}
\newcommand{\eea}{\end{eqnarray}}
\newcommand{\bml}{\begin{subequation}}
\newcommand{\eml}{\end{subequation}}

\input epsf

\begin{document}

\begin{opening}
\title{Cold and Warm Denaturation of Proteins}

\author{Guido Caldarelli \thanks{gcalda@pil.phys.uniroma1.it}}
\institute{INFM Sezione di Roma1 and Dipartimento di Fisica, Universit\`a "La
Sapienza", P.le A. Moro 2, 00185, Roma, Italy.}

\author{Paolo De Los Rios \thanks{Paolo.DeLosRios@ipt.unil.ch}}
\institute{Institut de Physique Th\'eorique, Universit\'e de Lausanne,
1015 Lausanne, Switzerland}

\begin{abstract}
We introduce a simplified protein model
where the water degrees of freedom
appear explicitly (although in an extremely simplified fashion).
Using this model we are able to recover both the warm and the
{\it cold} protein denaturation within a single framework,
while addressing important issues about the structure of
model proteins.
\end{abstract}

\keywords{Hydrophobic Forces, Proteins, Cold Denaturation}
\classification{PACS numbers}{}

\end{opening}

\section{Introduction}

Proteins are extremely complex structures: they are long heteropolymers
made of up to $20$ different amino-acids species,
each of them with its own chemical,
electrostatic and steric properties; the physiological solvent, an aqueous
solution, and its characteristics play a fundamental role both in the
dynamics and in the thermodynamics of folding.
It is therefore not surprising that only in recent times
statistical physicists have begun working
on this problem, mainly after the introduction of the so-called HP
model\cite{LD89+},
where the above mentioned richness has been reduced to a manageable
level.
In the HP model, proteins are modeled as self-avoiding polymers
on a lattice (two or three dimensional), greatly reducing
the number of accessible conformations\cite{Vanderzande}.
The chemical and electrostatic properties
of amino-acids have also been simplified: indeed,
it has been recognized since long
that the main force stabilizing the native conformations of
globular proteins is the {\it hydrophobicity} of non-polar
amino-acids\cite{Kauzmann59}.
Consequently, the important
properties of amino-acids are reduced to two: they are either polar
(ions or dipoles, labeled with P) or non-polar (H).

Hydrophobicity can be described as the tendency of hydrophobic molecules
to reduce as much as possible their surface of contact with water:
two hydrophobic molecules try to stick together in order
to hide from water their mutual surface of contact. Consequently,
hydrophobicity has been introduced in the HP  model as an effective
attractive interaction between H amino-acids. Then, the solvent
degrees of freedom can be neglected. Here we show that such a simplification
can be removed, and water can be taken into account, keeping the complexity
of the model at a still manageable level: the benefits are a
better description of the protein phenomenology (namely, {\it cold}
destabilization and eventually denaturation\cite{MPG90,Privalov1})
and some insights on the structure of the protein core.

In the last fifteen years there has been a growing body of evidence
for the so called {\it cold destabilisation} of proteins: the free energy
difference $\Delta F^N_D$ between denaturate and native
conformations of proteins has parabolic shape, with a maximum
at temperatures of the order of $15-25 ^0$C, or lower, implying that at lower
temperatures the native conformation is less and less stable.
In some cases, even the {\it cold denaturation}
of proteins has been obtained~\cite{Privalov2}.

There are at least two reasons to believe that a
good description of cold destabilisation and
denaturation is relevant to protein folding.

In order to describe protein folding with a simple model, it is
important to capture the essential physics of the process, at the temperatures
at which it takes place.
If the stability of native conformations of proteins begins to decrease
below $15-25 ^o$C, it is unlikely, at least {\it a priori},
that the physics responsible for such a behavior is not important around
the maximal stability temperature, in a range relevant for {\it in vivo}
protein folding.
A further reason to believe that a good model for protein folding should also
agree with the cold destabilisation phenomenology is that, actually,
there is no clear-cut distinction between the physics that stabilizes proteins,
and the one that destabilizes them. In both cases a re-analysis
of the concept of hydrophobicity and of hydrophobic hydration is necessary.

Already Frank and Evans~\cite{FE45+} identified the origin of hydrophobicity in
the partial ordering of water around non-polar molecules (such as, for
example, pentane, benzene and some amino-acids).
Water molecules tend to build ice-like cages around non-polar molecules.
Although a detailed analysis of these structures is,
to our knowledge, still lacking (actually recently some
better understanding and consensus are emerging\cite{Muller90,LG96,SHD98,SHD99}),
we can guess their energetic and entropic properties. Indeed,
water molecules forming these cages are highly hydrogen bonded, much as
in ice; consequently, their formation is energetically favorable with respect
to bulk liquid water.
Yet, the possible molecular arrangements in the cages are a small number
compared to all the disordered molecular conformations typical of liquid water.
The latter are
energetically unfavorable with respect to bulk water because
water molecules fail to form hydrogen bonds with hydrophobic amino-acids.
Therefore the free energy of formation of a cage ($F_{cage}-F_{no\;cage} =
\Delta F$) is a
balance between an enthalpy gain/loss and an entropy loss/gain:
ordered cages give an enthalpy gain ($\Delta H < 0$) and an entropy loss
($\Delta S <0$); the scenario is the opposite for disordered states.
All of the above arguments call for a model able to reproduce
(at least qualitatively) such a rich phenomenology.

\section{The HP-Water Model}

The model we propose here borrows two of the simplifications from the HP
model: proteins are still modeled as heteropolimers on a lattice,
made of just two different amino-acid species: polar (P) and non-polar (H).
Every site of the lattice that is not occupied by the
polymer is occupied by water (in general, by a group of water molecules that can
be arranged in $q$ states).

\subsection{Two-State Models for Water}

Water is described using the
Muller-Lee-Graziano (MLG) two-states model (Figure \ref{Fig1}a)~\cite{Muller90,LG96}.
The energy of each $H$ amino-acid depends on the states
of the water molecules it is in contact with (the water molecules in the hydration shell).
As mentioned above, hydration water can build ordered cages around the molecule,
that are energetically favorable with respect to the possible disordered
configurations, hence $E_{ds} > E_{os}$. Yet, the disordered configurations
outnumber the ordered ones: $q_{ds} > q_{os}$.
Water molecules that are not in contact with non-polar molecules ({\it bulk
} water) are described by a two-state model as well: water molecules can build
networks of hydrogen bonds that are energetically favorable with respect to disordered
configurations ($E_{db} > E_{ob}$) even if there are far more disordered configurations
than ordered ones ($q_{db} > q_{ob}$).
The above arguments hold separately for bulk and shell molecules.
In order to understand the order of the energies and of the degeneracies
we need to describe the effects of the transfer of water molecules
from the bulk to the hydration shell. Indeed, hydrogen bonds between hydration shell water molecules
on the average are stronger than hydrogen bonds in the bulk ($E_{ob} > E_{os}$);
conversely, the number of hydrogen bonded configurations in the hydration shell
is smaller than in the bulk ($q_{ob} > q_{os}$). Actually, the two inequalities
are mutually consistent: the greater the number of equivalent configurations, the higher
the probability to switch from one to the other; therefore, the persistence time of the bulk
hydrogen bonds is shorter than the persistence time in the hydration shell,
whence the energy inequality. The ordered bulk orientations that do not contribute
to ordered shell configurations can be counted among the disordered shell configurations,
that are therefore more abundant than the disordered bulk configurations. Moreover, such configurations
are energetically less favorable than bulk disordered configurations because every time a water
molecule points one of its bonding directions toward the non-polar molecule, it loses energy:
therefore $E_{ds} > E_{db}$ and $q_{ds} > q_{db}$.
Such a hand-waving picture has been recently confirmed by Silverstein {\it et al.},
who derived the double two-state MLG model
using a molecular model of the water-amino acid system~\cite{SHD99}.

The simple MLG model is extremely effective in describing the thermodynamics
of solvation of hydrocarbons (that, as recalled above, are strongly hydrophobic: indeed
the residues of hydrophobic amino-acids are essentially hydrocarbons, {{\it e.g.}, the residue of
leucine is an isopropyl group)~\cite{LG96}.
The degeneracies and energy differences can be fitted to experiments, in order
to get the detailed values. The MLG model has six free parameters
(one degeneracy and one energy can be chosen as reference): too many for a
{\it simple} theoretical model. We therefore introduce the Bimodal model (BM):
as a simplifying assumption (see Figure
\ref{Fig1}b), we say
that out of the $q$ possible states of a water molecule, one can be singled out
to be a cage conformations (labeled $s=0$),
energetically favorable with energy $-J$
($J>0$), and the remaining $q-1$ ($s=1,...,q-1$)
states are energetically unfavorable with energy $K>0$
(they represent the disordered states of reduced hydrogen-bond coordination).
We stress that the term {\it (un)favorable}
is always with respect to bulk liquid water. Bulk water molecules that are not
in contact with $H$ amino-acids
do not contribute to the energy.
Such a model is much simpler than the MLG one, yet it bears qualitatively similar results, as we shall
show.

\subsection{The Model}

As we mentioned above, we model proteins as polymers on a lattice. Monomers can be either
hydrophobic or polar. Sites that are not occupied by any amino-acid, are occupied
by water. The energy of the model is given by the energy of the water sites.
Every water site is occupied by {\it some} water molecules,
so that its available energy states should be given by a suitable
convolution of several MLG or BM models, as given in the previous section.
As a simplifying assumption, we describe the energetics of a water site
using a two-state model, choosing the bulk or shell parametrisation
depending on whether the site is in contact with a $H$ monomer or not.
$P$ amino-acids do not
interact with water so that their energy is always $0$: such a crude
approximation is made with the idea that hydrophobicity is the leading
effect stabilizing the native conformation of proteins. Some
better description of the water-P
interaction would be welcome, but
such ingredient is unnecessary for our present purposes.

Indeed, in the original formulation of the HP model, only interactions between
hydrophobic amino-acids where considered. Although this is clearly a strong
assumption, it has the advantage to keep the model as simple as possible and to
clearly address the effect of the sole hydrophobicity. On the other hand, the various
approximations of the model imply that the questions we can answer are somehow limited.
In this paper we look only at the thermodynamic behavior of proteins, and
at the simplest of the structural features, that is the segregation of hydrophobic
residues in the core of the protein. Other important problems can be addressed
already within the HP-model scheme, such as the relation between
sequences and structures, and in particular the designability of the latter.
We will tackle such issues in future works.

In what follows we will make explicit use of the BM model: formulas for the
MLG model are similar and can be easily derived.
Given a protein of $N$ amino-acids, with the sequence
$a_1,a_2,...,a_N$ ($a_i=P$ or $H$), the energy of
the protein is then
\begin{equation}
E = \sum_{<j,H>}{}(-J \delta_{s_j,0}+K(1-\delta_{s_j,0}))\;
\label{hamiltonian}
\end{equation}
where the sum is over the water sites that are nearest neighbors of some $H$
amino-acid.
Starting from (\ref{hamiltonian}) we can write the partition
function of the system as
\begin{equation}
Z_N = \sum_{C} Z_N(C)
\label{partition}
\end{equation}
where $Z_N(C)$ is the partition function associated to a single
conformation $C$:
\begin{equation}
Z_N(C) = q^{n_0(C)}\left((q-1) e^{-\beta K } + e^{\beta J} \right)^{n_1(C)}
\label{conf part func}
\end{equation}
and the dependence on the water degrees of freedom has been explicitly
calculated. $n_1(C)$ is the number of water sites nearest neighbors of
some $H$ amino-acid, $n_0$ is the number of water sites in the bulk or in contact only with
$P$ amino-acids.
We also keep the description of water as simple as possible, neglecting any interactions
between different water sites.

We deal with model proteins of length up to $N=17$ on the square lattice, and
compute the partition function, and all the thermodynamic quantities and
averages
by exact enumeration of the
$2155667$
different conformations.
We show the results for the particular sequence
$PHPPHPPHPHPPHPPHH$.
Such a sequence has been chosen to have a compact state with all
the hydrophobic amino-acids in the core. That such a state is the native one
(most stable and unique) is what we must check with the statistical
mechanical treatment.
We choose $J=1$ (actually,
both $K$ and the temperature $T$ can be normalized with respect to $J$),
$K=2$ and $q=10^5$ (a better determination of these values
could come from molecular dynamics and structural studies).
We take the Boltzmann constant $k_B=1$.

\section{Results}

In Figure \ref{Fig2} the specific heat $C_v$, and the average number of
monomer-monomer contacts, $n_c$, are shown. The low-temperature
peak in the specific
heat coincides with a jump of $n_c$: at lower temperatures
the protein is swollen, and maximizes the number of water-H contacts, in
agreement with cold denaturation.
The number of contacts, $n_c$, begins decreasing coinciding
with the high-temperature peak of the specific heat, that therefore coincides
with the usual
warm denaturation phenomenon.

Between $T_c$ and $T_w$ there is
a region where the most probable conformation is the one represented in the
inset of Figure \ref{Fig2}: as it can be seen, it is compact with a hydrophobic
core, out of reach for water (we also checked that this {\it native}
state is unique,
in that its Boltzmann weight is the largest above $T_c$).
We have analyzed the behavior
of different protein lengths and of different sequences, and we have
always found the same
qualitative behavior of $C_v$ and $n_c$.
Our model is therefore able to describe, within a single framework,
both cold and warm denaturation. Moreover, it shows a native state
with a mostly hydrophobic core.

Although the ratio between $T_w/T_c \sim 10$
in Figure \ref{Fig2} is unphysical (from experiments $T_w/T_c \sim 1.5$),
using the full MLG model it is possible to
come closer to real values: in Figure \ref{Fig3} the ratio is $T_w/T_c \sim 3$, and going toward more
and more refined models of water and of water/amino-acids interactions it is surely
possible to get physical values of the ratio. Of course the price to be paid is the larger number
of parameters to adjust. In this work we address the physical principles
responsible for the thermodynamic behavior of proteins on a broad range of
temperatures: we believe that the differences between the bimodal model and the
MLG model (and other possible more refined models) govern the details of the
behavior more than the essential features.

We next compare the free energy, enthalpy and entropy variations of
folding of our model with those from the literature\cite{MPG90,Privalov1}.
Indeed, such a comparison is a difficult one, since it is hard
to define what a denatured state is in our theoretical calculations, and even the
experiments have not yet been conclusive on such issue.
Therefore, as a simple approximation,
we consider as denaturate those conformations with at most
$4$ monomer-monomer contacts (a polymer of $17$ monomers over a square lattice
has at most $9$ monomer-monomer contacts).
The native state has $8$ monomer-monomer contacts.

In Figure \ref{Fig4} we show $F_{Denaturate}-F_{Native} = \Delta F^D_N$,
$\Delta H^D_N$ and $T \Delta S^D_N$. They coincide qualitatively
with the ones from experiments\cite{MPG90,Privalov1}.
We point out the presence of two temperatures
below and above which $\Delta F^D_N < 0$: the denatured state of our model
protein is more stable than the native state.
Between these two temperatures, instead,
$\Delta F^D_N > 0$, and the native state is the most stable.
In the same temperature range where $\Delta F^D_N > 0$, $\Delta H^D_N$ and
$T \Delta S^D_N$ have a strong temperature dependence: they even change sign, a
signature of the rich physics behind the water-protein system.
At high temperatures we find that both $\Delta H^D_N$ and $\Delta S^D_N$
saturate ($T \Delta S$ grows linearly, therefore $\Delta S$ saturates),
as experimentally observed\cite{Privalov1}.
Some particular care should be paid to the low temperature behavior of
$\Delta H^D_N$ and $T \Delta S^D_N$. Indeed, $\Delta H^D_N$ goes to a constant
value, which is consistent with a lower bound for the energies, and
$T \Delta S^D_N$ tends to $0$ with $T$. Experiments should be made
below $T_c$ to assess such a behavior (although a recent model suggests
such scenario~\cite{SHD98}).
We find therefore that our model reproduces qualitatively
the known calorimetric data of protein denaturation over a broad range of
temperatures.

\section{Effective Interaction}

\subsection{Two-Body Interactions}

The hydrophobic effect
is often modeled through attractive  effective $HH$ interactions.
Within our framework (and on a square lattice, for simplicity),
we consider a system of two $H$ amino-acids in solution and we compare
the partition function of the system when the two amino-acids
are in contact,
\begin{equation}
Z_c = q^2 (e^{\beta J} + (q-1) e^{-\beta K})^6\;\;,
\end{equation}
with the one when the distance between the two
amino-acids is infinite,
\begin{equation}
Z_0 = (e^{\beta J} + (q-1) e^{-\beta K})^8.
\end{equation}
The effective
interaction is defined as
\begin{equation}
\epsilon_{HH} = T \ln\frac{Z_c}{Z_0} = 2 T \ln\frac{q}{e^{\beta J} + (q-1) e^{-\beta K}}\;\;
\end{equation}
($\epsilon_{HH}$
is positive if attractive, with this definition).
The $T\to\infty$ limit is
\begin{equation}
\epsilon_{HH}(T\to\infty) = 2K - \frac{2}{q}(K+J)
\end{equation}
and is attractive for large values of $q$: it is the usual
hydrophobic effective interaction. Yet, the $T=0$ limit
is $\epsilon_{HH}=-2J$, repulsive. A meaningful effective interaction
should at least include such a temperature dependence.
Actually, the strong temperature dependence of $\epsilon_{HH}$ is not the
only limitation to a definition of an attractive $HH$ interaction.
Indeed, such an interaction can be meaningful only for amino-acids
surrounded by water molecules, but it cannot be defined in the core
of proteins, where water is absent. As a consequence, in the
absence of some true interactions between amino-acids, the hydrophobic
interaction alone is not able to favor thermodynamically
the native state against different compact states obtained by reordering
only the core of the protein. As an example, the two conformations
in Figure \ref{Fig5}, corresponding to the sequence $PPHPPPPPPPPPHHHP$
have the same probability to occur in our model, since they hide and expose to
water the same number of H amino-acids.

\subsection{Many-Body Interactions}

When a protein is folding, its amino-acids
find an {\it ever changing} environment that depends on water and
on the other amino-acids. Even the
reliability of two-body effective interactions vs. many-body ones is an open
issue still to be settled.
In our model we can compute some many body effective interactions.
First, we observe that next-nearest-neighbor interactions are equal
to nearest-neighbor interactions, $\epsilon_{nnn} = \epsilon_{HH}$, since there are again
six water sites in contact with the hydrophobic molecules.
Then we consider three $H$ particles in an {\it angle} configurations (see Figure \ref{Fig6}a).
The effective energy can be computed as
\begin{equation}
\epsilon_{HHH} = 5 T \ln\frac{q}{e^{\beta J} + (q-1) e^{-\beta K}} \ne 2 \epsilon_{HH} + \epsilon_{nnn}
\,\, .
\end{equation}
Already  this simple situation shows that integrating away the solvent degrees of freedom
cannot be reduced to simple two-body effective interactions: the intrinsic three-body contribution to
the free energy amounts to roughly $20\%$ of the total.
Many-body effects can be pronounced also if the corner amino-acid is polar ((see Figure \ref{Fig6}b).
In that case the effective energy is
\begin{equation}
\epsilon_{HPH} = 7 T \ln\frac{q}{e^{\beta J} + (q-1) e^{-\beta K}} \ne  2 \epsilon_{HP} + \epsilon_{nnn}
\end{equation}
with $\epsilon_{HP} = T \ln[(q/( e^{\beta J} + (q-1) e^{-\beta K})]$.
Actually, the non additivity and more generally the context-dependence of water mediated
effective interactions has also been recently pointed out~\cite{sBBMPP98}.

\subsection{Validity of Effective Interactions}

This model suggests that it is improper to define
interactions of hydrophobic origin inside proteins, and that
the detailed structure of the cores of proteins should be stabilized
by other mechanisms. Recently, many methods have been devised to derive effective potentials
able to stabilize protein structures: some of them \cite{Skolnik} are of a statistical
nature and have been shown to be intrinsically flawed \cite{VD98}; some other methods that have
a more rigorous physical basis have also been proposed \cite{SMMB+}.
Still, no matter the physical soundness of the method,
the presence of intrinsic many-body effective potentials casts a shadow over any
simplistic two-body description of amino-acid/amino-acid interactions.

Moreover, effective interactions can be safely defined whenever they substitute
some non-changing environment \cite{BM01}.
It is therefore intrinsically difficult to define
effective potentials of some general validity between amino-acids:
our model points out such a problem for hydrophobic interactions.

\section{Conclusions}

In conclusion, we have introduced a model of proteins in water
that is able to reproduce the known features of proteins
(namely, cold destabilisation and warm denaturation,
a native state with a mostly hydrophobic core,
and the correct free energy, enthalpy and entropy
of folding).
We also checked our results for different protein lengths,
sequences, parameter values and even implementing the full MLG model
for the description of water. Although some details may change,
the overall behavior is consistent and robust. Moreover, lattice models
are intended to be only qualitatively instructive, whereas a quantitative
description can be given only by more detailed off-lattice models.

Our model is a first step in an interpolation between fully microscopic
models, where water is dealt with at a molecular level, and fully effective models,
where water is accounted for by effective potentials between amino-acids.
As we have shown, the reliability of simple two-body, context-independent effective potentials
is, at least as a matter of principle, questionable (very recently, Park {\it et al.}
proposed at least a distinction between surface and core two-body potentials~\cite{PVD00}):
within our scheme such a problem emerges clearly. Although our model is an extremely
simplified one, we believe that moving toward more realistic descriptions
of the water/amino-acid system would complicate the structure of the effective potentials,
rather than simplifying it, and that therefore a much better understanding is needed.

\begin{acknowledgements}
The authors thank E.I. Shakhnovich and S. Vaiana for useful suggestions.
This work has been partially supported by the Swiss National Research Fundation
under grant FNRS 21-61397.00 and by the European Network Contract
FMRXCT980183.

\end{acknowledgements}

\newpage

\begin{figure}
\epsfysize= 7 truecm
\begin{center}
\mbox{\epsfbox{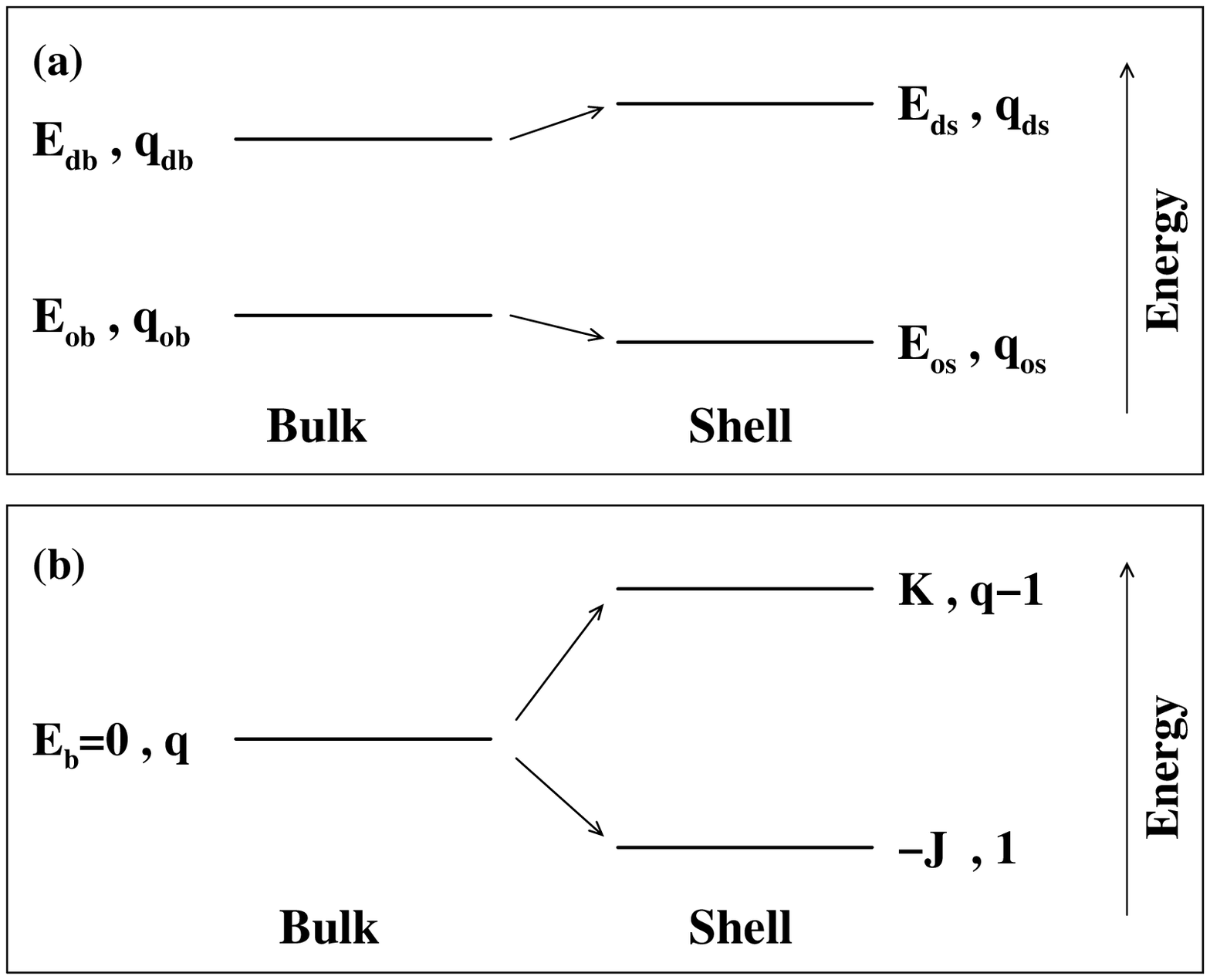}}
\end{center}  
\caption{Bimodal effective models. Panel (a): MLG model, with bimodal energy
distributions both for bulk and shell water molecules. The lower levels
represent ordered group of water molecules, the higher levels disordered ones.
The order of energies and of degeneracies, as obtained from experiments,
is $E_{ds} > E_{db} > E_{ob} > E_{os}$ and $q_{ds} > q_{db} > q_{ob} > q_{os}$
($ds$ = disordered shell, $os$ = ordered shell, $db$ = disordered bulk, $ob$ = ordered
bulk). Panel (b): the simplified bimodal energy distribution, with just two
free parameters, $K$ and $q$, since we can take $J$ as energy scale.}
\label{Fig1}
\end{figure}

\begin{figure}
\epsfysize= 7 truecm
\begin{center}
\mbox{\epsfbox{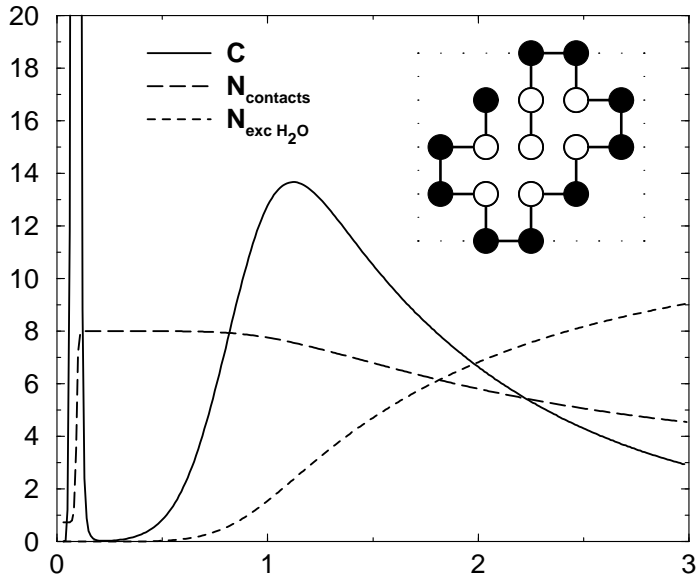}}
\end{center}  
\caption{Specific heat, monomer-monomer contacts and number
of water sites in an excited state for the protein shown in the inset;
$K=2$, $J=1$, $q=1000$.}
\label{Fig2}
\end{figure}

\begin{figure}
\epsfysize= 7 truecm
\begin{center}
\mbox{\epsfbox{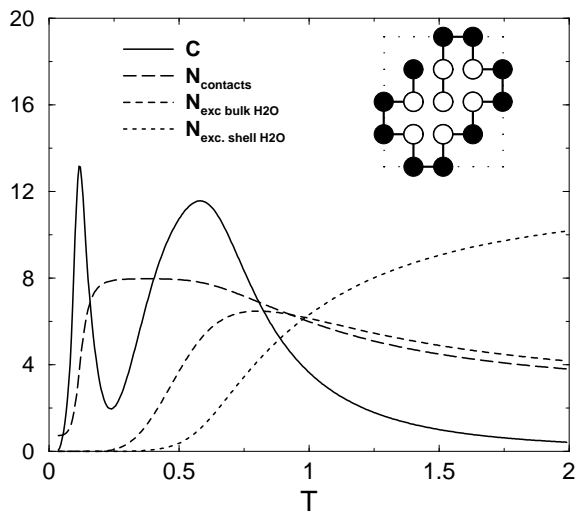}}
\end{center}  
\caption{Same as in Figure 2, for a full implementation of the MLG model, with
$E_{os}=-1.4$, $E_{ds}=1.8$, $E_{ob}=-1$, $E_{db}=1$, $q_{os}=1$, $q_{ds}=999$,
$q_{ob}=50$, $q_{db}=950$.}
\label{Fig3}
\end{figure}

\begin{figure}
\epsfysize= 7 truecm
\begin{center}
\mbox{\epsfbox{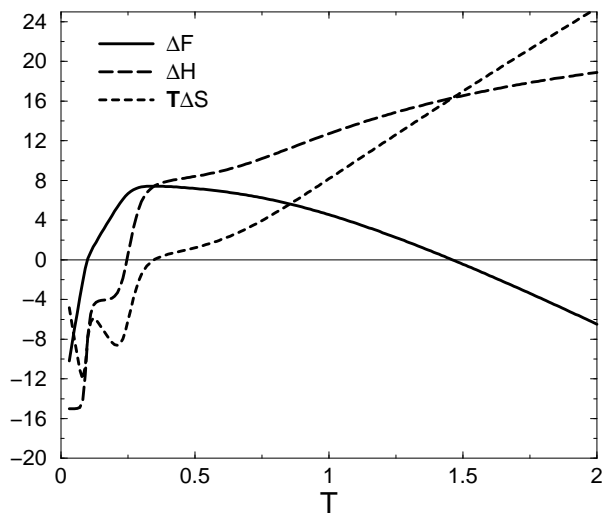}}
\end{center}  
\caption{Free energy, enthalpy and entropy (times $T$)
differences between denatured conformations and the native one (shown in the
inset of Figure 2), for the same parameter values as in
Fig.$3$. Since $T \Delta S$ grows linearly at high temperatures,
$\Delta S$ saturates.}
\label{Fig4}
\end{figure}

\begin{figure}
\epsfysize= 7 truecm
\begin{center}
\mbox{\epsfbox{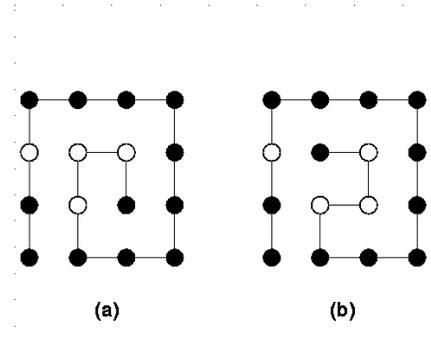}}
\end{center}  
\caption{Two different conformations of the same sequence
differing only for a reorganization of the core amino-acids.}
\label{Fig5}
\end{figure}

\begin{figure}
\epsfysize= 7 truecm
\begin{center}
\mbox{\epsfbox{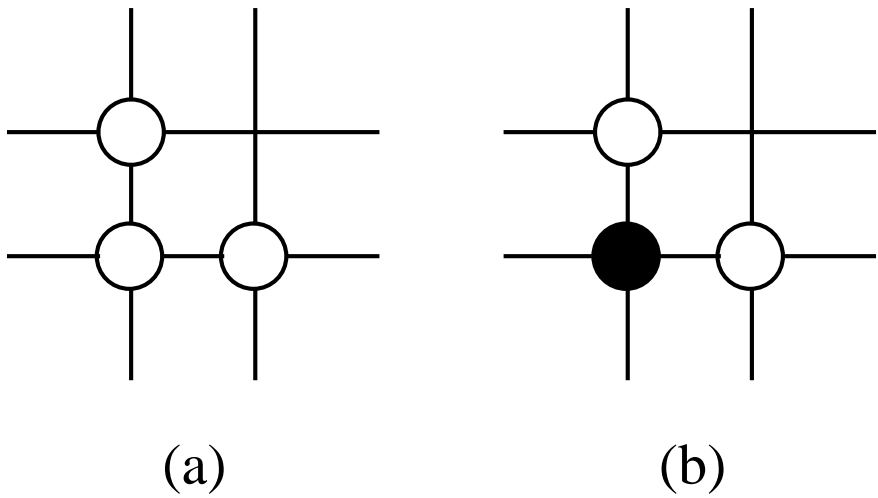}}
\end{center}  
\caption{Angle configurations used to compute the three-body effective energies:
a) three $H$ particles and b) two $H$ and one $P$ particles.}
\label{Fig6}
\end{figure}

\end{document}